# N/MEMS Biosensors:
# An Introduction

Vinayak Pachkawade

VPACHKAWADE Research Center,
Nagpur, India
vinayak@vpachkawade.com

**Abstract.** In the 21st century, biosensors have gathered much wider attention than ever before, irrespective of the technology that promises to bring them forward. With the recent COVID-19 outbreak, the concern and efforts to restore global health and well-being are rising at an unprecedented rate. A requirement to develop precise, fast, point-of-care, reliable, easily disposable/reproducible and low-cost diagnostic tools have ascended. Biosensors form a primary element of hand-held medical kits, tools, products, and/or instruments. They have a very wide range of applications such as nearby environmental checks, detecting the onset of a disease, food quality, drug discovery, medicine dose control, and many more. This chapter explains how Nano/Micro-Electro-Mechanical Systems (N/MEMS) can be enabling technology towards a sustainable, scalable, ultra-miniaturized, easy-to-use, energy efficient, and integrated bio/chemical sensing system. This study provides a deeper insight into the fundamentals, recent advances, and potential end applications of N/MEMS sensors and integrated systems to detect and measure the concentration of biological and/or chemical analytes. Transduction principle/s, materials, efficient designs including readout technique, and sensor performance are explained. This is followed by a discussion on how N/MEMS biosensors continue to evolve. The challenges and possible opportunities are also discussed.

**Keywords:** Biosensors, N/MEMS, Health & Wellbeing, BioMEMS, Applications.

## 1    Introduction

Biosensors are playing a crucial role in today's biomedical science and technology [1]. Biosensors allow the sensitive and precise detection of a range of biological or chemical contaminants [1–4][5]. It is a device that measures biological or chemical reactions and generates an output signal in proportion to the concentration of a target analyte in the reaction. Typical elements of a biosensor are shown in Figure 1. It consists of the following elements, 1) A small inlet/channel ($\mu$fluidic) to collect/navigate the sample/solution to the sensing system. A sample may be in the form of a solid, liquid, gas, or combination thereof. A sample in solid form may contain dust, dirt, soot, or smoke particles. Samples in the gaseous form include air with a mixture of oxygen, nitrogen, carbon dioxide, or other hazardous gas elements. A liquid sample may include bodily fluids, for example, nasal swabs, saliva, blood, semen, sweat, urine, etc. Analyte/s in the samples are a target substance to be detected. For example, particulate matter (PM) is an analyte whose concentration in the surrounding air is to be measured. Triglyceride is an analyte whose concentration in the blood is to be measured. In clinical applications, analytes of interest are glucose, vitamins, hemoglobin, molecules, proteins, amino acids, urea, bodily gases, toxins, specific biomarkers (to detect chronic disease), living cells, and pathogens (viruses and bacteria). 2) Next element is called bio receptors (see Figure 1). Bio receptors are a group of molecules that are deposited/immobilized (as a thin film/layer) onto the surface of a transducer. Bio receptors interact with the sample/solution to precisely recognize the target analyte/antigen. Such interaction between the bio receptor molecules and target analyte is termed biorecognition. The output of a biorecognition event may result in a change in mass, light, color, temperature, pH level, etc. 3) a sensor converts this form of energy into a measurable signal in either optical or electrical form (conductance/impedance, charge/current, potential/voltage, frequency/phase, etc.). Here, we use the words transducer and sensor interchangeably. 4) A sustaining electronics in the bio device/module can process the sensor output by further amplification, filtering, analog-to-digital (A/D) conversion, and microprocessor/memory storage. Eventually, the module shows the reading on the computer or an embedded display in real time. A wireless data transmission may also be added to the unit.

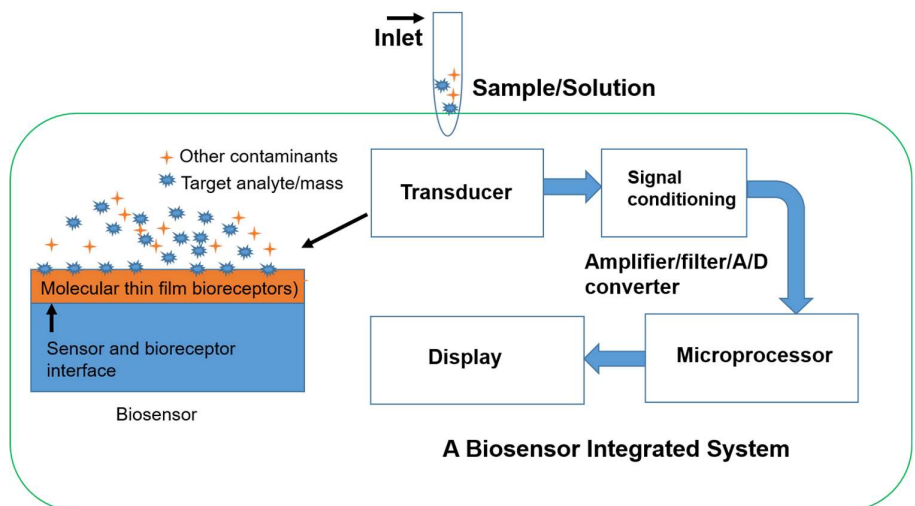

**Fig. 1.** A typical bio/chemical sensing platform. Bio receptors enable the selective detection of a target analyte in the sample. A transducer's job is to convert the biorecognition process into measurable output. Such measurable output is then correlated to the detection and concentration of a target analyte.

Figure 2 shows the information on the potential areas of applications of biosensors. Biosensors are widely being deployed for environmental check (to detect hazardous gas elements), to detect the onset of the existing/future disease, monitoring the quality of food and beverages at places where these are stored, on farms to inspect soil quality, in medicine dose control, and many other.

**Potential application areas of biosensors**

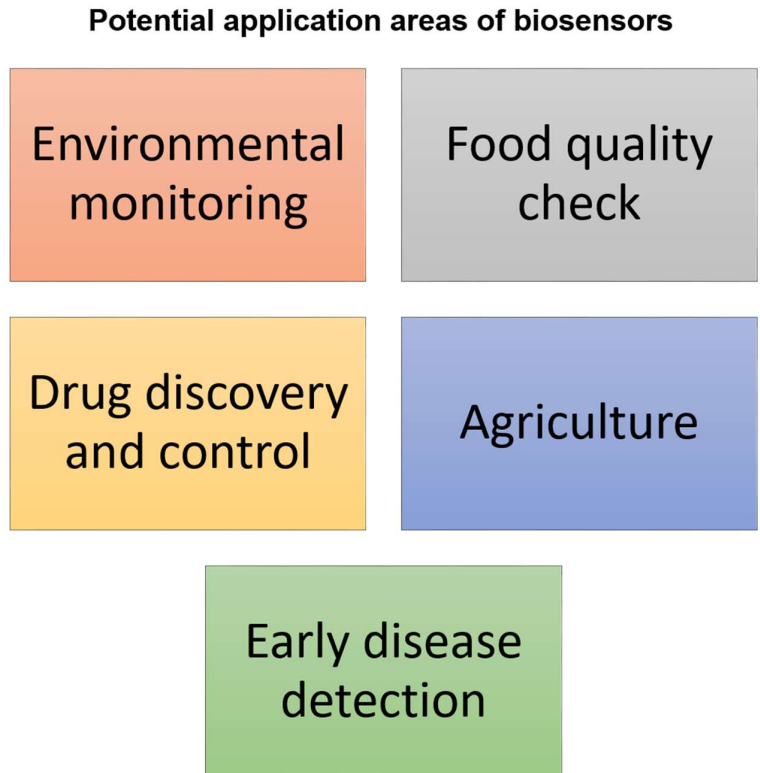

**Fig. 2.** A diagram showing potential application areas where biosensors are used. There are many other areas where biosensors are primarily used.

## 2    Characteristics of a biosensor

In the biosensing development platform, the following parameters are of the most importance. Researchers are continuously finding ways to improve the following features (see Figure 3).

**Sensitivity/Resolution**: It is defined as the smallest possible change in the physical/biological/chemical properties of a transducer that can be resolved after the biomass/target analyte/s are adsorbed/absorbed by the sensor surface (i.e., after the reaction occurs). Sensitivity (also expressed in %) can be given as the ratio of change in the sensor output per unit change in the input. It is also called the detection limit. A mass sensor, for example, can show resolution up to fg/ml to record the concentration of analyte traces in a liquid sample. Measuring a lowest possible concentration of a specific analyte/s (for example, antigen) can be associated with an underlying medical condition for which doctors can prescribe a test.

**Precision:** Precision is an important feature of a biosensor. Precision/Selectivity/Specificity indicates the ability of a bio receptor thin film molecule to detect/recognize a specific analyte within a mixture of other contaminants in a sample under test (see Figure 1). An example of precision is the contact of an antigen with the antibody. Typically, antibodies act as bio receptors and are immobilized on the surface of the transducer. A solution/sample that contains the antigen is exposed to the transducer, where antibodies should interact specifically with the antigens. To develop a biosensor, precision is the key characteristic.

**Stability**: The stability of a sensor is the characteristic to avoid responding to changes in the environmental conditions (i.e. temperature, humidity, vibrations, pressure, etc.), thereby preventing a false output. To address this issue, a provision is made in the embedded electronics to compensate for environmental drifts in the sensor output. Such common-mode signals can be canceled using a differential sensor readout.

**Reproducibility:** Reproducibility is the capability of the biosensor module to produce the same output when the experiment/test is repeated. It implies that the sensor provides an average value close to the correct value when a sample is measured several times. Reproducible sensor output indicates the high consistency and robustness of a biosensor.

**Noise floor:** This parameter provides the baseline above which a sensor output can reliably be detected/measured. Noise in the sensing system stems from i) the intrinsic noise in the sensing element and ii) electronics. Therefore, a sensor is designed to reduce the overall noise. Lower the noise floor, better the sensor signal-to-noise ratio (S/N), or say sensitivity/resolution.

**Linearity and Dynamic range:** A sensor can show linear changes in the output per unit change in the input. The output of a sensor should produce a linear response to different concentrations being measured. This characteristic is also often represented by the % nonlinearity in the sensor output. Linear dynamic range (expressed in dB) is the ratio of the maximum to the minimum sensor output.

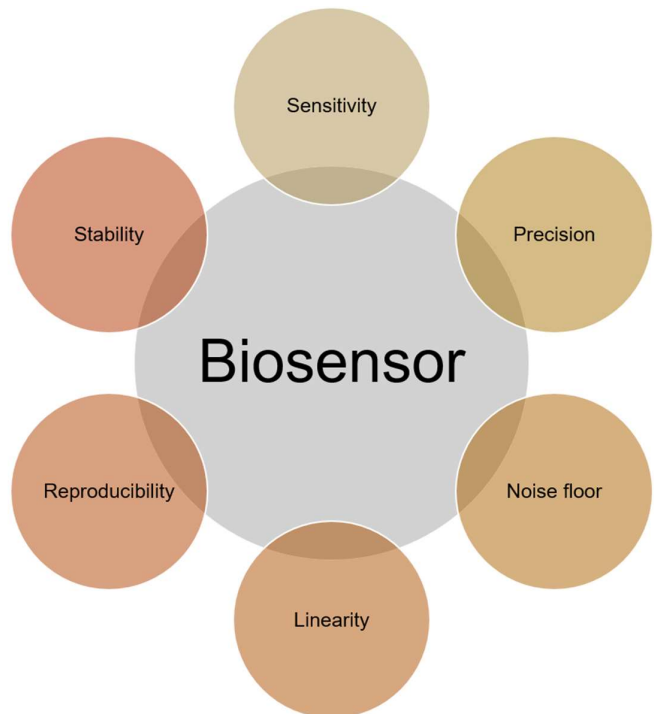

**Fig. 3.** Performance parameters in biosensor development.

# 3    N/MEMS Biosensor (BioMEMS)

N/MEMS feature scalability, ultra-high sensitivity, energy efficiency, and compatibility to easily integrate with microelectronics [6,7][8]. N/MEMS offers outstanding precision in the detection and measurement of ultra-small elements. This advantage comes with the miniaturization N/MEMS offers. For example, a nominal mass of N/MEMS can scale down to nanograms or even smaller. It is therefore possible to detect and measure things at $10^{-15}$, $10^{-18}$, or even lower scale. Miniaturization also means a small footprint, weight, energy, and internet of sensors (IoS), all at reduced system-level cost. With N/MEMS, it is possible to precisely detect the target analyte and measure its concentration. This is achieved through precisely recording even the minutest shift in the sensor output [4,9,10]. Transduction principles such as piezo-electricity, piezo-resistivity, electro-static, electro-magnetic, electro-thermal, optical, etc. are popularly used in the development of the N/MEMS enabled bio/chemical sensors.

### 3.1 Mass accumulation

N/MEMS devices are operated by the principle of gravimetric sensing [7,11]. Mass accumulation is characterized by the adsorption/absorption (biorecognition) of the mass of a particle or target element in a biological/chemical sample. N/MEMS transducer is a mechanical cantilever/tuning-fork/plate/disk/bar/membrane. This transducer is set in a static or a dynamic mode (vibrate at a particular frequency called a natural resonant frequency). Figure 4 shows a schematic illustration of a N/MEMS transducer that can be used as a primary element in a typical N/MEMS biosensing platform. As seen in Figure 4, a transducer (cantilever in this case) in a static form is used for sensing purposes. A transducer surface is functionalized to attract specific molecules. The functionalization takes place by depositing/loading nanoparticles (for example gold nanoparticles) onto the surface of a cantilever. After this, target molecules are attached to the nanoparticles. Such mass accumulation leads to the gradient in i) the surface stress (σ) (see Figure 4 (a)), ii) surface strain, and iii) position/displacement of a cantilever (see Figure 4 (b)). In other words, the transducer is actuated by the component of the force exerted by the bio/chemical reaction that occurs at the surface of a transducer.

For the sensing part, a set of resistors (forming a Wheatstone bridge electronic circuit) is embedded at the other end of a cantilever, where it is fixed/anchored (see Figure 4). At this end, the stress is maximum. Another viable method of sensing is to monitor a change in the parallel plate capacitance at the free end of a cantilever. At this end, displacement is maximum. A bottom electrode placed beneath the free end of the cantilever can be used to detect the capacitive gradient when the cantilever moves.

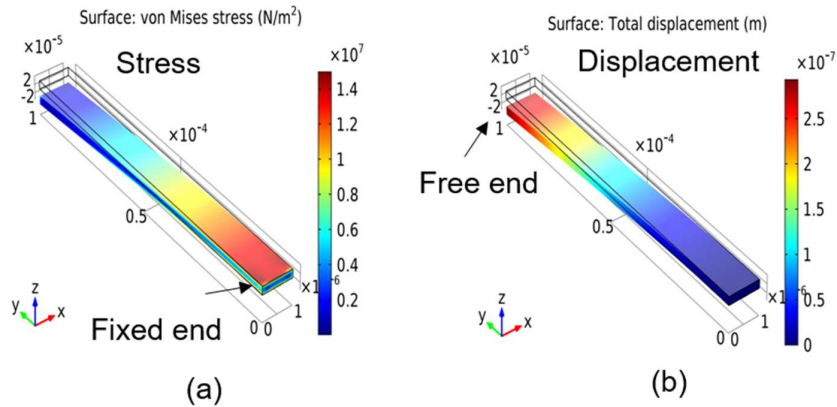

**Fig. 4.** A static N/MEMS transducer that can be used to sense and quantify target bio/chemical contaminants in a sample. Part (a) shows a stress profile of a structure that can be used to use piezo-resistive transduction. Part (b) shows a displacement profile of a structure that can be used to use capacitive transduction.

### 3.2 Materials

Usually, silicon/polysilicon is used to construct such a transducer. In recent years, piezoelectric materials (for example, aluminum nitride (AlN) and quartz) have become a popular choice in realizing several potential applications in BioMEMS [4]. However, materials such as silicon, piezo, etc. are suited only for non-invasive applications, a procedure that does not require inserting an instrument through the skin or into a body. Such materials can be a part of N/MEMS-enabled biosensors. These sensors can then be used for monitoring environmental parameters, lab-on-chip, laboratory tests, wearable gadgets, etc. Given the biocompatibility issue, materials such as Polydimethylsiloxane (PDMS) and other polymers are also widely used for biosensing [12]. Another application of PDMS is that a transducer can be embedded into the $\mu$channels made by PDMS for lab-on-chip fluid testing applications.

### 3.3 N/MEMS resonators for ultra-precise bio/chemical sensing

N/MEMS resonators are one of the most promising candidates to develop a wide range of applications in medical diagnostic and instrumentation [1,4,8,13–16]. N/MEMS resonators offer extremely high-quality factors ($Q$) and high frequency. These two parameters result in exceptionally high parametric sensitivity. Additionally, the N/MEMS resonant sensing platform provides good output stability. Due to their highly precise output, N/MEMS resonators are extremely useful to sense environmental, physical, biological, and chemical quantities. They can detect multiple analytes in parallel, and offer immunity toward a false output. Being a primary component in many commercial applications, N/MEMS resonators are today a popular choice to develop low-cost, miniaturized, reliable, reproducible, and precise sensing solutions. Possible areas of applications are point-of-care/remote monitoring of target parameters in the environment, agriculture, medicine, health, and well-being, thus democratizing sensing across the globe. N/MEMS resonators operate on the principle of gravimetric sensing. The N/MEMS mass sensor is reported to detect particle concentration as low as attogram ($ag$) or even zeptogram ($zg$). Figure 5 shows a representative example of a N/MEMS resonator that can be used to detect and quantify the concentration of a target contaminant in a biological/chemical sample. Upon adsorption/absorption of a target trace/particle, a shift in the resonant frequency of a transducer can be accurately recorded (in real-time). These shifts can be correlated to the concentration of the contaminant/s. The essential assumption is that adsorption/absorption of a mass, $\Delta m$ is much smaller than the nominal mass, $M$ of the resonant transducer ($\Delta m << M$).

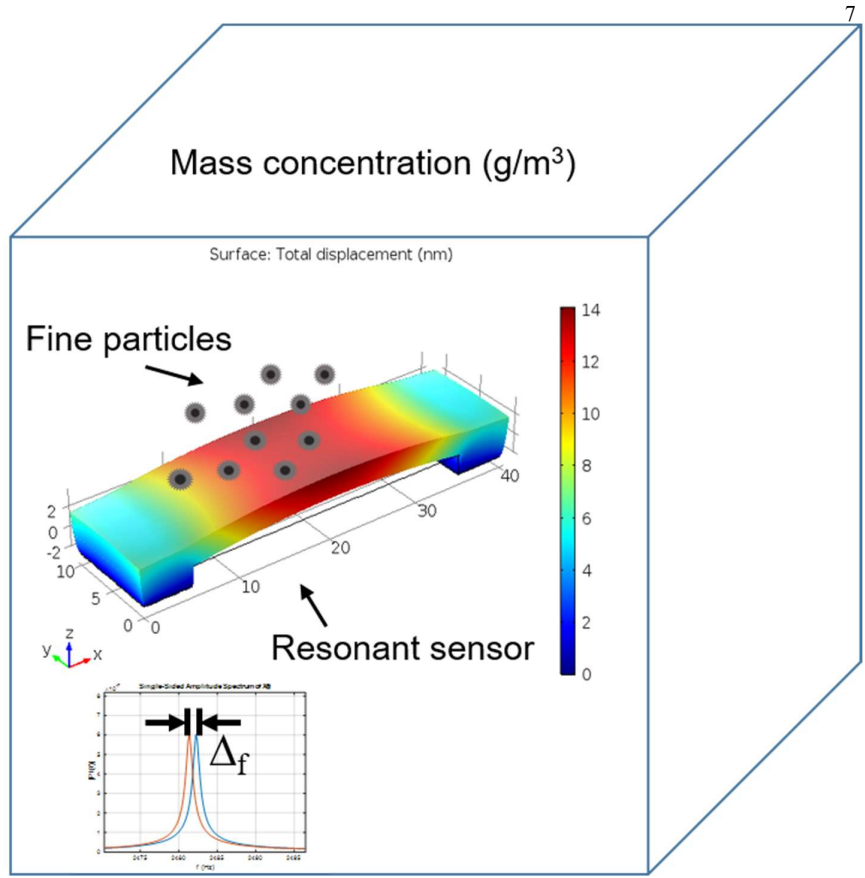

**Fig. 5.** A representative example schematic of a resonant mass sensor for particle detection and concentration measurement of a target analyte. $\Delta f$ indicates the shift in the resonant frequency, i.e. transducer output.

### 3.4    Multiple analyte detection using N/MEMS

Figure 6 shows a platform parallel processing using N/MEMS sensors. As seen, $C_1$ to $C_n$ are the cantilevers arranged as an array. These sensors are attached to a common base. Each of the cantilevers is coated/immobilized/functionalized with a *specific* bio receptor thin film to attract the target molecules/particles. Such a sensing platform is highly useful for the fast testing of multiple parameters in bio/chemical samples. Due to simultaneous testing and processing, the overall cost is low. Such a platform is compact and efficient. However, a specific coating of *n* number of sensing units in an array is required. Such a requirement may be a trade-off with the advantages in terms of cost and time.

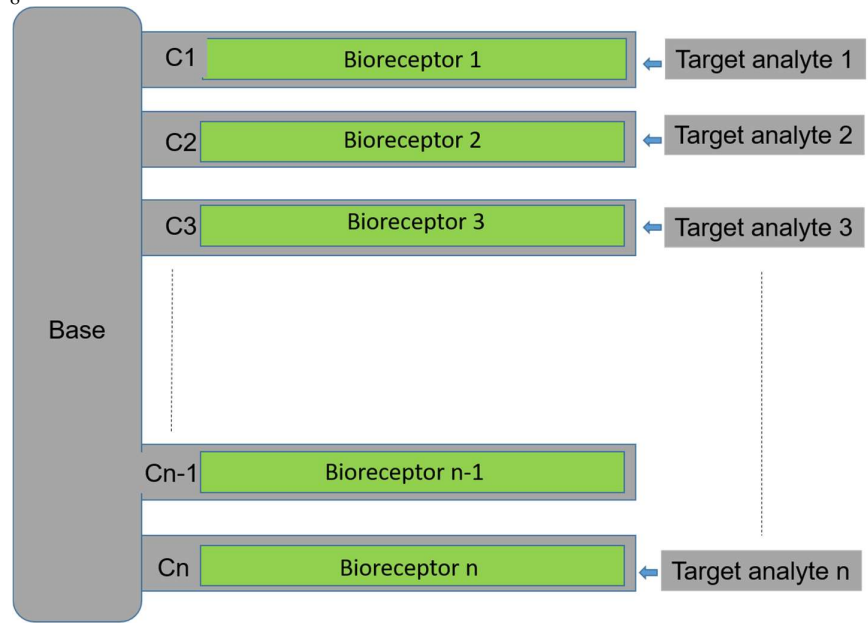

**Fig. 6.** An array of N/MEMS sensors to specifically detect, and measure the concentration of a target analyte in a sample. A parallel test and processing of multiple analytes in the sample are useful in terms of cost and time.

## 4 Summary, Outlook, and Conclusion

This chapter introduced the basics of biosensors and how N/MEMS sensors can be used in several bio/chemical applications. The chapter started by providing an overview of a typical N/MEMS-enabled bio/chemical sensing platform. The N/MEMS sensors can be used to detect and precisely quantify several biological and chemical contaminants. This scientific area requires expertise, experience, and collaborations across the multi-domains (mechanics, optics, biology, chemistry, physics, and microelectronics). A static mode N/MEMS can be used to monitor the surface stress, strain, or displacement when the mass of a target particle/analyte is attached to the sensor. An electro/optical sensing mechanism or a combination thereof can then be used to read and correlate the sensor output to the presence and concentration of the target element/s in a sample. A dynamic (resonant) mode N/NEMS is also used as a high-sensitivity transducer in biological and chemical applications. Here, a transducer element is set to vibrate at the designed frequency. When the mass of a target particle/analyte is adsorbed/absorbed on the chemically functionalized surface of a transducer, a change in the effective mass of the resonant transducer results in a change in the frequency.

N/MEMS-enabled hand-held biosensor modules can be employed for many applications. These applications are healthcare, surgery, drug discovery/dose control, recognition of pollutants, detection of micro-organisms responsible for causing a disease, and markers to indicate disease in bodily liquids/fluids (blood, urine, saliva, semen, sweat, etc). Typical applications of BioMEMS also include sensors measuring intravascular blood pressure, therapeutics applications (e.g., drug delivery actuators, disease monitors), pacemakers, and defibrillators. Biosensors can be implantable or wearable. These are also part of sensing systems for real-time monitoring of several body parameters such as pulse rate, blood pressure, oxygen, and neurological activities.